\documentclass{IEEEtran}
\usepackage[ruled,vlined]{algorithm2e}
\usepackage{graphicx,
	psfrag,
	epsfig,
	epstopdf,
	amsthm,
	amssymb,
	url,
	subcaption,
	balance,
	enumerate,
	color,
    url,
	setspace,
	tikz,
	pgfplots
}
\usepackage{amsmath}
\usepackage[nospace,noadjust]{cite}
\usepackage{pbox}
\usepackage{geometry}
\usepackage{multirow}
\usepackage{adjustbox}
\usepackage{array}
\usepackage{booktabs}
\usepackage{float}
\usepackage{tikz}
\usepackage[noend]{algpseudocode}
\usepackage{tabularx}

\newcommand{\cref}[1]{Constraint~\ref{#1}}

\newcommand{\ignore}[1]{}
\newgeometry{left=1.2cm,right=1.2cm,bottom=1.5cm,top=1.5cm}
\definecolor{dark}{rgb}{0.0, 0.5, 0.0}

\begin{document}


\title{Towards Zero-Trust 6GC: A Software Defined Perimeter Approach with Dynamic Moving Target Defense Mechanism}

\author{
	\IEEEauthorblockN{Zeyad Abdelhay \IEEEauthorrefmark{1}, Yahuza Bello\IEEEauthorrefmark{1}, and
	Ahmed Refaey\IEEEauthorrefmark{1}\IEEEauthorrefmark{2}}

				\IEEEauthorblockA{\IEEEauthorrefmark{1}University of Guelph, Ontario, Canada}\\
	\IEEEauthorblockA{\IEEEauthorrefmark{2}Western University, London, Ontario, Canada}}

\maketitle
\begin{abstract}

The upcoming Sixth Generation (6G) network is projected to grapple with a range of security concerns, encompassing access control, authentication, secure connections among 6G Core (6GC) entities, and trustworthiness. Classical Virtual Private Networks (VPNs), extensively deployed in Evolved Packet Core (EPC) network infrastructure, are notoriously susceptible to a variety of attacks, including man-in-the-middle incursions, Domain Name System (DNS) hijacking, Denial of Service (DoS) attacks, port scanning, and persistent unauthorized access attempts. This paper introduces the concept of Software Defined Perimeter (SDP) as an innovative solution, providing an alternative to VPNs with the goal of fostering a secure zero-trust milieu within the 6G Core networks. We capitalize on the SDP controller-based authentication and authorization mechanisms to secure the EPC network's control and data plane functions, conceiving an architecture that is expansible to the 6G network. Further, we augment the SDP zero-trust capabilities via the incorporation of a dynamic component, the Moving Target Defense (MTD). This enhances the network's resilience against attacks targeting traditionally static network environments established via VPNs. Following rigorous testbed analysis, our proposed framework manifests superior resilience against DoS and port scanning attacks when juxtaposed with traditional VPN methodologies.

\end{abstract}

\section{Introduction}

\begin{figure*}[ht!]
    \includegraphics[width=1\textwidth, height=14cm]{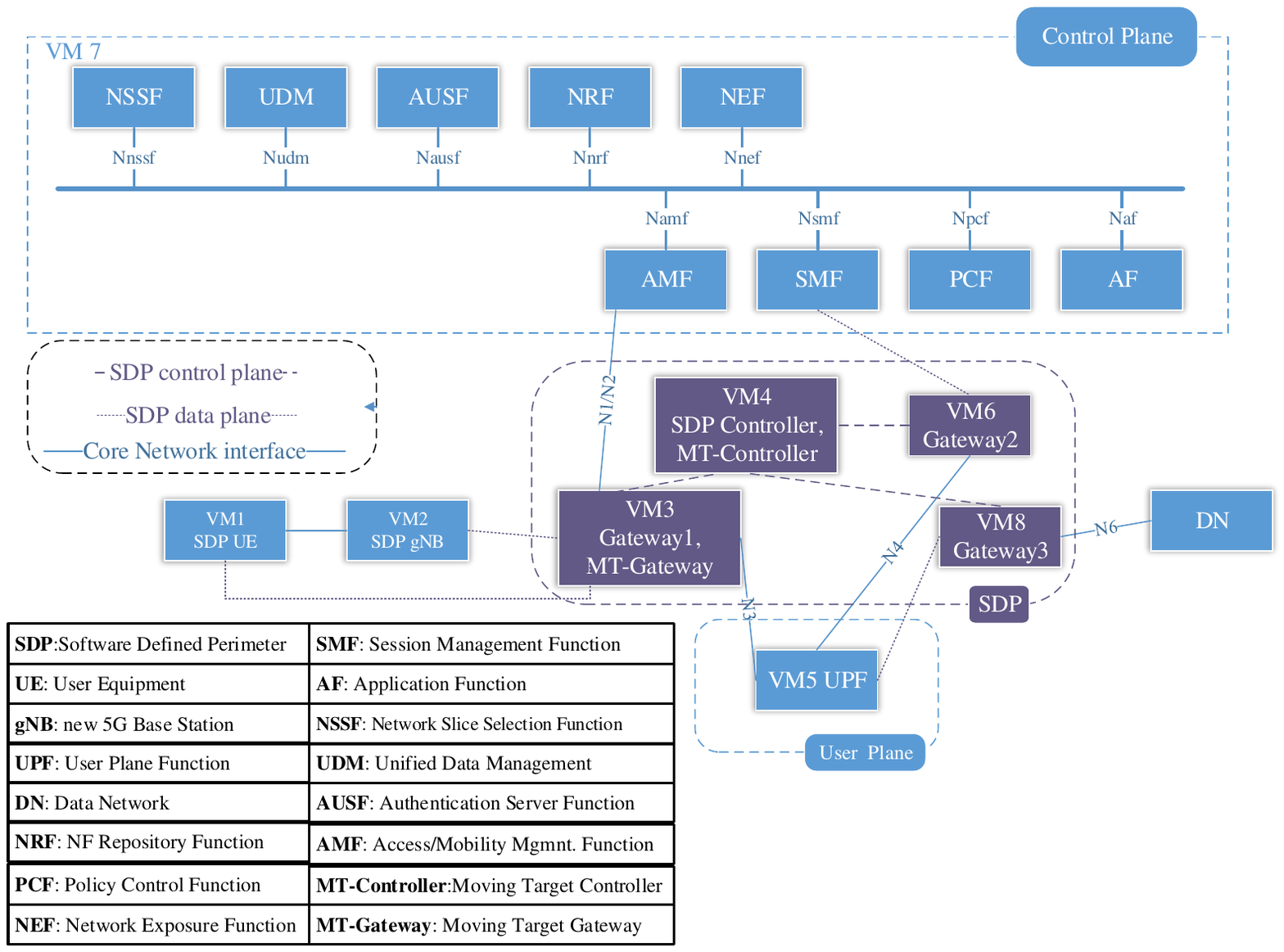}
     \caption{Proposed SDP-based EPC network architecture and deployment model with MTD capability.}
    \label{fig:5G-SDP}
\end{figure*}

\IEEEPARstart{T}{he} evolution of mobile networks led to network performance, capacity, and data rate advancements. Sixth Generation (6G) networks will support services like Internet of Everything (IoE), extended reality, smart grids, and Intelligent Transportation Systems (ITS) \cite{9397776}. However, securing 6G networks while transitioning from 5G poses significant challenges\cite{9524814} inherited by vulnerabilities in 5G networks like Artificial Intelligence advancements that affected the reliability of RSA cryptosystems, and 5G network vendors advocating for the implementation of Virtual Private Networks (VPNs) for unauthorized access prevention. Despite their merits, VPNs are susceptible to a range of attacks, including man-in-the-middle and DoS. To address these limitations, a more stringent zero-trust security architecture is necessitated, paving the way for solutions such as the Software Defined Perimeter (SDP) \cite{9729545} \cite{x17}.

Indeed, SDP implements Zero Trust principles and has demonstrated resilience against various cyber attacks in previous research studies \cite{sdn-sdp,nfv_sdp} \cite{x16}. By adopting SDP, the proposed architecture and deployment model for 6G networks ensures a zero-trust environment through dynamic firewall configurations and SDP-based authentication and authorization. This deployment model can be leveraged to 6G networks replacing VPN-based deployments that are currently proposed by major cellular network vendors \cite{wpaper}.

To substantiate the effectiveness of this proposed model, a lab-scale testbed was utilized for its implementation. When juxtaposed with OpenVPN in this controlled environment, the experimental results clearly indicated the enhanced capability of the model, with respect to SDP's superiority in repelling several attacks. However, SDP remains vulnerable to attackers who can exploit its static nature once the communication link is established to gain insights into the network. Consequently, this might violate the zero trust principles where attackers can acquire the time and knowledge needed to persistently launch successive attacks.


In an effort to mitigate this plausible security risk and strengthen the zero trust tenets upheld by SDP, we have integrated Moving Target Defense (MTD) within our proposed architecture and deployment model. This strategic addition dynamically modifies the attack surface. MTD, in essence, utilizes Network Address Shuffling (NAS) to trigger regular address changes, enhancing network security. By introducing an element of uncertainty and reducing the timeframe for potential intruders to probe and initiate attacks, NAS serves as a powerful deterrent \cite{x15}.



Therefore, the main contributions of this work can be summarized as follows:

\begin{itemize}
    \item Propose an architecture and deployment model that integrates SDP for Evolved Packet Core (EPC) networks, establishing a zero-trust environment using dynamic firewall configurations and SDP-based authentication and authorization. This approach, tested in a lab-scale testbed, exhibited superior performance over VPN which is currently recommended in 5GC networks. 
    \item Recognizing possible vulnerabilities in the potential static situation of SDP that could be exploited by attackers, we further bolstered the model's security by integrating MTD. This dynamically alters the attack surface, enhancing the robustness of the model. 
    \item Propose the integration of Network Address Shuffling within the MTD framework which introduces an element of unpredictability to the model. By frequently altering network addresses, NAS significantly enhances network protection, shortening the window for potential attacks and adding an extra layer of security deterrent. 
\end{itemize}

The subsequent sections provide foundational insights into VPN deployment within EPC networks, SDP as a zero-trust model, and MTD. We subsequently introduce a unified architecture fusing SDP and MTD, devised to fortify network-level security within EPC's landscape as a use case. Given the anticipated similarity of EPC Core in future 6G networks, this proposed architecture holds substantial merit for impending 6G deployments. Following that, we delve into an evaluation of the testbed's performance, and the discourse culminates with key conclusions and future research directions.



\section{Foundational Overview}

\begin{figure*}[h]
    \centering
    \includegraphics[width=0.8\textwidth, height=13cm]{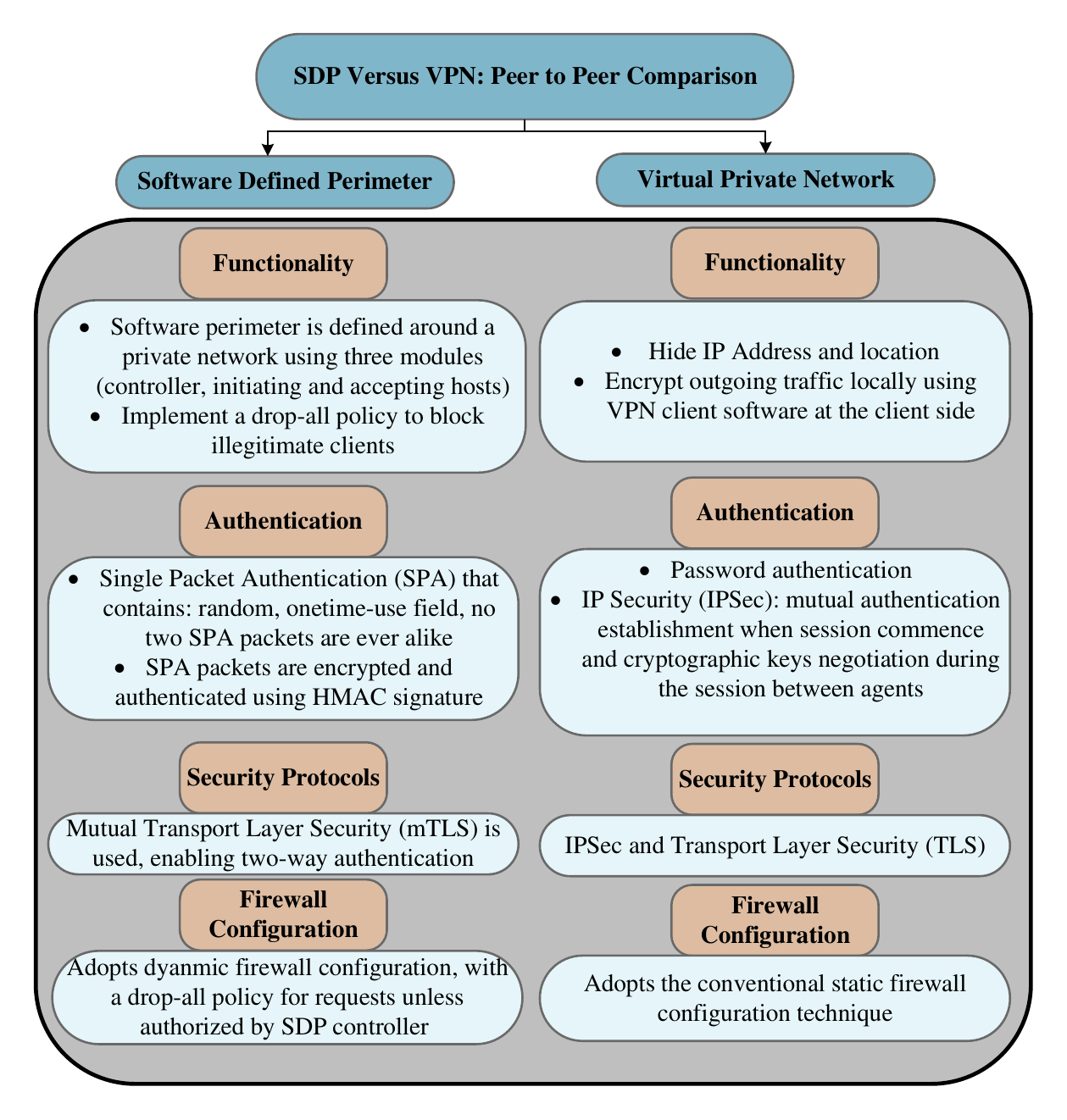}
    \caption{Software defined perimeter versus virtual private network.}
    \label{fig:SDPVPNCOMP}
\end{figure*}

In this section, we explain prior research landscape relevant to our work. We start with a general description of the EPC network entities which are followed by a basic description of VPN, SDP, and MTD, and the section is completed with 
Figure~\ref{fig:SDPVPNCOMP} and Table I, which presents a more detailed explanation of the SDP framework in comparison with VPN in terms of functionality, authentication mechanism adopted, security protocols used, firewall configuration, and some security-related features\cite{x15}\cite{x14}. 


\subsection{Evolved Packet Core (EPC) Network Entities}
Wireless mobile networks consist of the Radio Access Network (RAN) and the EPC network. The EPC RAN consists of User equipment (UEs) and gNBs (Next Generation Base station). According to the ETSI reference model, the EPC network adopts a microservice-like architecture consisting of various network functions. These different network functions as presented in Figure~\ref{fig:5G-SDP} vary in functionality ranging from Access and Mobility Management Function (AMF) to Network Slice Selection Function (NSSF). Moreover, the CN adopts the Control and User Plane Separation of the CN nodes (CUPS) whereby the control plane functionality and the user plane functionality (i.e., as shown in Figure~\ref{fig:5G-SDP}) are decoupled.
For interested readers, refer to \cite{5GC2} for a more detailed explanation of all EPC Network Functions.


EPC is still adopted globally for wireless networks, and it has not yet replaced completely by 5GC. With the rapid continued developments towards 6G, it is anticipated that future 6G networks will have similarities to EPC deployment model. The virtual nature of EPC network functions makes it prone to attacks and security threats originating from the RAN and Core network communication. Especially user identity security, which might lead to attackers impersonating existing UE within the network, granting them access to vital network virtualized functions. Rendering VPN deployment useless and highly questionable, presenting a gap that VPN does not fulfill and highlighting the value of a software-based approach like SDP.

\subsection{Virtual Private Network (VPN)}

VPN is widely used for secure communication in EPC network deployments \cite{wpaper}. It forms a private link between an endpoint and a network resource. This link can exist within an insecure network. VPN employs an authentication-based method. This method allows access between the endpoint and the network resource. VPN functions within that network. It sets up a boundary around network resources.
Within this approach, traffic is encrypted, and resources are strictly shared among authorized users who are allowed access to the inside perimeter. This establishes an entry point to the network, acting as a prime security measure.
There are many topologies for VPN such as Peer-to-Peer connections, client-to-server connections, and site-to-site connections \cite{7314859}. Among these topologies, the most widely adopted one is the client-to-server VPN, in which a secure tunnel is established between a VPN client and server. Enabling secure transfer of encrypted data between VPN client and server. Since VPN is usually between authorized set of users, a strong access protocol is required to secure the network. For interested readers, refer to \cite{7314859} for an in-depth understanding of different protocols. 

OpenVPN is a layer 2 and layer 3 tunneling VPN protocol, which is open-source software for community usage. It implements a VPN client and VPN server and utilizes the OpenSSL for encryption purposes. OpenVPN uses a pre-shared and certificate-based method to authenticate authorized VPN clients and servers. For comparison purposes, we adopt OpenVPN within the EPC network, which will be later explained in section IV.

\begin{table*}
\centering
    \begin{center}
    
    \begin{tabular}{ | m{10em} | m{20em}| m{20em} |}
    \hline
        \textbf{Security Features} & \textbf{SDP} & \textbf{VPN} \\
    \hline
        Access Control & SDP provides security for on-premises servers and remote users simultaneously. Allowing remote access to the on-premises servers. & When VPN gateways are deployed to protect on-premises servers, VPN clients are not granted remote access to these servers. \\
    \hline
        Capital and operational cost & SDP's components' dynamic nature enables faster setup and management. Admins can protect multiple servers using the controller within seconds. & VPN’s static nature requires more administration and maintenance. Admins need to constantly update complex configurations to grant/revoke access to different resources. \\
    \hline
        Attack mitigation & SDP accepting hosts can be deployed in a different location from protected servers. This approach hides the location even from authorized clients. Efficiently securing SDP against DoS attacks. & VPN server deployment within the network enables attackers to identify the location. If the VPN connection drops briefly, an attacker can identify the real IP of a host. \\
    \hline
    \end{tabular}
    \caption{Security features of SDP and VPN}
    \end{center}
    \label{tab:SDPVPNSEC}
\end{table*}



\subsection{Software Defined Perimeter (SDP)}

The SDP framework follows a zero-trust model proposed by the National Institute of Standards and Technology (NIST). NIST provides required guidelines to design a Zero Trust Architecture (ZTA) to secure enterprise network services. Adopting this zero-trust model, the SDP framework consists of a controller, an Initiating Host (IH), and an Accepting Host (AH) \cite{10197716}. 

The controller handles authentication and authorization of all hosts (i.e., IHs and AHs) and provides access protocol for all available services. Within the controller module, authentication keys (i.e., Single Packet Authentication (SPA)) for all hosts are generated and stored in a database (MySQL is adopted for this purpose) and later distributed to the hosts. Any host requesting access to any protected service must present these SPA keys and only upon successful authentication by the controller can they be authorized to access them through the designated SDP gateway. 

The AH module is responsible for enforcing rules set by the controller (i.e., blocking all users from accessing services except those authorized by the controller). This is achieved by adopting a drop-all policy in the AH module where all requests are dropped by default before successful authentication. With this model, only the IHs with valid SPA keys can gain access to protected services within the network. It is worth mentioning that the AH module adopts a dynamic firewall configuration with the Firewall KNock OPerator (FWKNOP).

\subsection{Moving Target Defense (MTD)}
MTD emerges as a highly effective security method offering robust defense against reconnaissance activities and network mapping. This approach employs a dynamic strategy by continuously modifying the network characteristics, mitigating the inherently static nature of traditional systems that inadvertently grant attackers an advantage in terms of exploiting potential paths and vulnerabilities \cite{9644442}.

In the context of this work, we have opted to implement MTD utilizing Random Host Mutation (RHM) technique. Where the MTD Controller is responsible for assigning random virtual IP addresses from a pool of unassigned IPs to machines hosting VNFs while retaining the original IPs and establishing mappings accordingly. To ensure a high mutation rate and maintain unpredictability, the controller imposes a brief lifespan for these virtual addresses \cite{10.1145/3411496.3421223}.

MTD framework comprises a controller and a gateway. The primary function of the controller (MT-Controller) is conducting network address scanning in order to identify service hosts available within the network. The controller then establishes a virtual IP (vIP), from the pool of unassigned IPs within the network, for the service to be addressed publicly while maintaining a virtual-to-real IP mapping. Access to the protected services is restricted to hosts utilizing virtual IP (vIP), the controller will deny any requests utilizing real IP (rIP).

The gateway (MT-Gateway) performs address translation from virtual IP (vIP) to real IP (rIP) and vice versa, enabling effective routing of traffic to requested services. Additionally, the MT-Controller utilizes connection tracking to ensure seamless connection continuation, even after the shuffling process that leads to a change in the assigned vIP.

\section{Proposed Deployment Model and Architecture}
In a standard Zero-Trust Architecture (ZTA) per NIST guidelines, any access request must undergo authentication and continuous monitoring throughout the access duration. Hence, deploying ZTA components within a network is essential to uphold the zero-trust principle. The abstract ZTA components outlined by NIST include the Policy Decision Point (PDP) and Policy Enforcement Point (PEP). The PDP, composed of the Policy Engine (PE) and Policy Administrator (PA), generates session authentication credentials for specific resources and manages the establishment and termination of secure communication paths. Conversely, the PEP enforces policies set by the PDP, monitoring, enabling, and terminating access requests within the network. When applying the ZTA concept to Software Defined Perimeter (SDP), the SDP controller assumes the responsibilities of the PDP, while the SDP gateway handles the duties of the PEP. Therefore, the proposed combined zero-trust architecture as shown in Figure~\ref{fig:5G-SDP} consists of the network components (only the available implemented components were included, which are AMF+SMF, UE+gNB, and UPF) and SDP components (i.e., controller and gateways). As depicted in  Figure~\ref{fig:5G-SDP}, gateway 1 is deployed in front of the CN to blacken it to unauthorized users. Similarly, the SDP controller is deployed behind gateway 1 to shield it from unauthorized users. Within the CN, two more gateways (i.e., gateway 2 and 3) were deployed to provide a zero-trust environment where all entities (i.e., in this context AMF+SMF and UPF) involved require verification and authentication through the SDP controller prior to having access to one another. 

With this model, all UEs and gNBs must first authenticate to the controller before access to the CN is granted. Ensuring that only a positive cross-check of the SPA certificate for any UE+gNB will be considered by gateway 1 for access authorization. According to the ETSI reference model, the UE and the gNB communicate with the AMF through designated interfaces, N1 and N2 respectively. In the same fashion, gNB communicates with the UPF through N3 interface, and UPF-SMF communications are carried out through the N4 interface. The UPF provides the link to DN through N6 interface. Note that the proposed model adopts these interfaces without any modifications as shown in Figure~\ref{fig:5G-SDP}. It is worth mentioning that all control and data traffic of the network is routed through dedicated gateways as illustrated in Figure~\ref{fig:5G-SDP}.

Algorithm 1 presents a Pseudo-code for UE+gNB client authentication and authorization procedure to access AMF+SMF and UPF servers via SDP. The default setting of the SDP gateway is to discard all packets unless verified and authorized by the controller. First, the UE+gNB client requests access to AMF+SMF server by sending an access request (i.e., the SPA packet) to SDP gateway 1. If SPA packet is valid, SDP gateway 1 forwards the packet to the controller to be verified and to set the authorization rules and update the UE+gNB credentials. As soon as UE+gNB client is verified by the controller, it updates its credentials and informs SDP gateway 1 services authorized for UE+gNB client (i.e., AMF+SMF and UPF). SDP gateway 1 then updates the firewall rule settings to allow UE+gNB client's traffic to be forwarded to the intended services within a certain time period, $t$. As long as time $t$ does not expire, SDP gateway 1 maintains that rule and forwards UE+gNB client's to the services. As soon as the $t$ expires, SDP gateway 1 removes the firewall rule and returns to the default setting. 

Note that Algorithm 1 considers the scenario where UE+gNB server tries to access AMF+SMF and UPF server via SDP. The same approach applies within the EPC when AMF+SMF server tries to access UPF server.

\begin{algorithm}[h]

  \KwOut{Secure connection to AMF+SMF and UPF servers established via SDP\;}
  UE+gNB client requests access to AMF+SMF and UPF server via SDP gateway 1\;
  
  \eIf{the SPA packet sent is valid}{
   SDP gateway 1 forwards the SPA packet to the controller for verification and credential update;\;
   
   \eIf{UE+gNB client is verified by the SDP controller}{
   SDP controller creates a secure mutual TLS tunnel between itself and UE+gNB to update credentials;\;
   
   SDP controller informs SDP gateway 1 about AMF+SMF and UPF servers authorized to be accessed by UE+gNB via another secure mutual TLS;\;  
  
   SDP gateway 1 updates firewall rules to allow traffic forwarding for $t$ duration among UE+gNB, AMF+SMF and UPF servers;\;
   
   \eIf{UE+gNB requests access to either AMF+SMF or UPF server}{
   \eIf{the time SDP gateway 1 configured its firewall rule for UE+gNB $<$ $t$}{
   SDP gateway 1 forwards UE+gNB traffic to AMF+SMF and UPF servers;\;}{Remove firewall rule configured for UE+gNB;\;}

   }{Discard the packets;\;}
 
   }{Discard the packets;\;}
   
   }{Discard the packets;\;}
   
 \caption{UE+gNB client authentication within the proposed EPC-SDP architecture}
\end{algorithm}


This proposed architecture and deployment model provides a zero-trust environment within the EPC network entities and between the RAN and the EPC network. Additionally, the segregation between the control plane and data plane in the proposed model and the implementation of mTLS (mutual transport layer) protocol helps protect the connection between the SDP controller and the SDP gateways from any sniffing or eavesdropping activity, providing a much-needed security framework for the network, dropping any incoming traffic from intruders and unidentified hosts.

Additionally, as part of the proposed deployment model, MTD is seamlessly integrated, augmenting the network's security with an additional layer. This integration involves deployment of the MT-Gateway in front of the CN within gateway 1. MT-Gateway's role is to exclusively permit authorized users who utilize the virtual IP (vIP) to access services while effectively blocking any traffic using real IP (rIP) or an expired vIP. This deployment establishes a clear segregation between the virtual and real network addresses, thus enhancing the security posture of the overall system. In a similar manner, deployment of the MT-Controller takes place on the machine hosting the SDP controller, situated behind the gateway. The MT-Controller assumes the responsibility of assigning virtual IPs to both the services and gateway, while concurrently maintaining a mapping of virtual to real (V2R) and real to virtual (R2V) addresses. Additionally, the MT-Controller diligently keeps track of all open connections within the network.

The MT-Gateway plays a pivotal role in facilitating access to services requested by UEs and gNBs, utilizing virtual IP (vIP) addresses. Upon receiving a request, MT-Gateway thoroughly examines the packet to verify that the vIP aligns with a corresponding real IP (rIP). Subsequently, the request is forwarded to the SDP controller for authentication and validation of UE+gNB access. Once requester's authenticity is confirmed, the MT-Gateway modifies the source IP address to the vIP of the gateway, enabling communication with the requested service. In response to the request, acknowledgments are transmitted using gateway's vIP as the destination. Consequently, the MT-Gateway adjusts source and destination IP addresses to match the rIPs of the requester and gateway. This establishes an open connection, allowing the flow of traffic between the requester and the service.
As the MT-Controller triggers a timeout event and initiates another instance of IP mutation. Throughout this process, the MT-Controller imposes connection tracking to effectively manage any ongoing connections, mitigating the risk of service interruption.

\section{Testbed and Performance Evaluation}

In this section, we provide our testbed specifications and details of the complete implementation. We then move forward to discuss the results of the performance evaluations.

\subsection{Testbed Specification}
\subsubsection{Adopted open source project}
The implementation of the testbed consists of three open-source projects: Open5gs for the EPC and EPC NR, Waverlay Lab's SDP project for the SDP framework, and OpenVPN project for VPN. Open5gs provides implementation of AMF+SMF, UPF, simulated UE, and gNB on VMs, which we adopt for our testbed scenario as shown in Figure~\ref{fig:5G-SDP}. Waverly Lab's SDP project provides the SDP client module, SDP controller module and SDP gateway module, which we implemented on VMs as shown in Figure~\ref{fig:5G-SDP} and the OpenVPN project provides the VPN client, VPN server, and Certificate Authorization (CA) server implemented on VMs as well. Note that the VPN client is implemented on VM 1 and VM 2 while the VPN server is implemented on VM 5 and VM 7 as depicted in Figure~\ref{fig:5G-SDP}. The CA server is implemented on a separate dedicated server for importing and signing certificate requests of both VPN client and server.

\subsubsection{Specs of the host servers and VMs}
\begin{itemize}\item The server hosting the VMs is running Linux Ubuntu 18.04 LTS Bionic Beaver, VirtualBox 6.2 and is dedicated to serving as the CA server for the VPN with Intel Core i7-7700 CPU @3.60GHz 8, 12 GB DDR4 and 1 Gbps full duplex ethernet card. \end{itemize} 
\begin{itemize}\item VMs 1, 2, 5 and 7 are running Linux Ubuntu 20.04 with 1 vProcessor and 1 GB of RAM dedicated for hosting RAN (SDP UE and SDP gNB) and CN (user plane and control plane functionalities) as illustrated in Figure~\ref{fig:5G-SDP}. \end{itemize} 
\begin{itemize}\item VMs 3, 4, 6 and 8 are running Linux Ubuntu 16.04 Xenial with 1 vProcessor and  1 GB of RAM for the SDP framework (i.e., SDP/MT controllers and SDP/MT gateways) as shown in Figure~\ref{fig:5G-SDP}. \end{itemize}

\subsubsection{Configuration of the SDP project}
According to testbed setting with respect to proposed model, the UE+gNB server will attempt to access AMF+SMF and UPF servers through the SDP gateway on ports 44 and 45 respectively. The gateway is configured to forward traffic received on those two ports (only legitimate clients' traffic after positive verification is accepted) to the services (i.e., AMF+SMF server and UPF server) on ports 7777 and 8888 respectively. For VPN settings, the UE+gNB server serves as the VPN client while both AMF+SMF and UPF serve as VPN servers waiting to accept requests from configured clients.

\subsubsection{Attack scenario and evaluation metrics}
To evaluate the performance of the proposed model, we compare effectiveness of SDP framework with that of VPN in terms of four metrics: time required to initialize the components of both SDP and VPN, the overheads introduced to EPC network by both SDP and VPN, the average throughput and Round-Trip Time (RTT) of both SDP and VPN under DoS attack while up and running. A single instance of port scanning attack was performed under both SDP and VPN. This is because it is sufficient to demonstrate whether open ports can be detected in any testbed environment. The performance of any proven solution is often constrained by the availability of computing resources (usually in terms of CPU and memory). To this end, we analyze both CPU and memory usage of SDP and VPN components implemented in the network. We then perform a heartbleed attack with SDP. Heartbleed attack is a well-known vulnerability in OpenSSL library (version  1.0.1 to 1.0.1f), which allows an adversary to steal data such as usernames, passwords, private keys, TLS session keys, etc from the victim's server. Note that OpenVPN implementation in our testbed uses Openssl version 1.1.1 and thus, not vulnerable to heartbleed attack.

\subsection{Results and Discussions}

Table II presents the results for four metrics. The SDP framework exhibits a slightly higher initialization time of 3.2637 seconds compared to VPN's 2.9537 seconds. This is due to the startup and connection process of SDP's framework three components. Despite shorter initialization time of VPN, SDP provides a stricter authentication and authorization approach. Regarding the induced overhead in the network, VPN shows lower overhead compared to SDP. SDP's overhead arises from the controller and gateway components, while VPN authentication involves the CA server, which can be vulnerable to attacks. SDP proves its superiority over VPN. Under a DoS attack, SDP achieves higher throughput and lower RTT. VPN's performance is affected by the DoS attack, resulting in lower throughput and higher RTT. It is important to note, this test case was conducted for 30 seconds, a duration frequently utilized in the literature \cite{ning2012design}, across three consecutive attempts, under both SDP and VPN. Hence, reported values given in Table II are the averaged outcomes.

\begin{table}[ht!]
\begin{center}
\caption{Performance Evaluation Results}
\begin{tabular}{ | m{5em} | m{5em}| m{5em}| m{5em}| }
  \hline\hline
  \multicolumn{4}{c}{VPN} \\
  \hline\hline
 Initialization time (s) & Induced overhead (s) & RTT (ms) & Average throughput (Mbytes/s)\\
  \hline
 2.9537 & 0.0513 & 3.54 & 50 \\
\hline
\multicolumn{4}{|c|}{\begin{minipage}{0.45\textwidth}
      \includegraphics[width=1\textwidth]{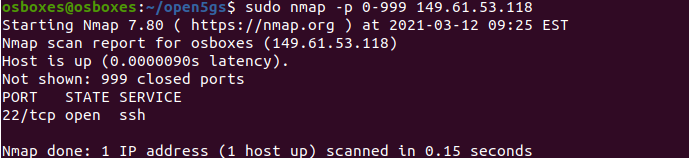}
    \end{minipage}} \\
    \hline\hline
  \multicolumn{4}{c}{SDP} \\
 \hline\hline
 Initialization time (s) & Induced overhead (s) & RTT (ms) & Average throughput (Mbytes/s)\\  
 \hline
  3.2637 & 0.0936 & 1.02 & 498  \\
  \hline
  \multicolumn{4}{|c|}{\begin{minipage}{0.45\textwidth}
      \includegraphics[width=1\textwidth]{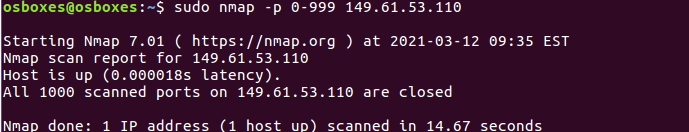}
    \end{minipage}} \\
 \hline
\end{tabular}
\end{center}
\label{table:PE}
\end{table}

\begin{figure}[h]
    \centering
    \includegraphics[width=.5\textwidth]{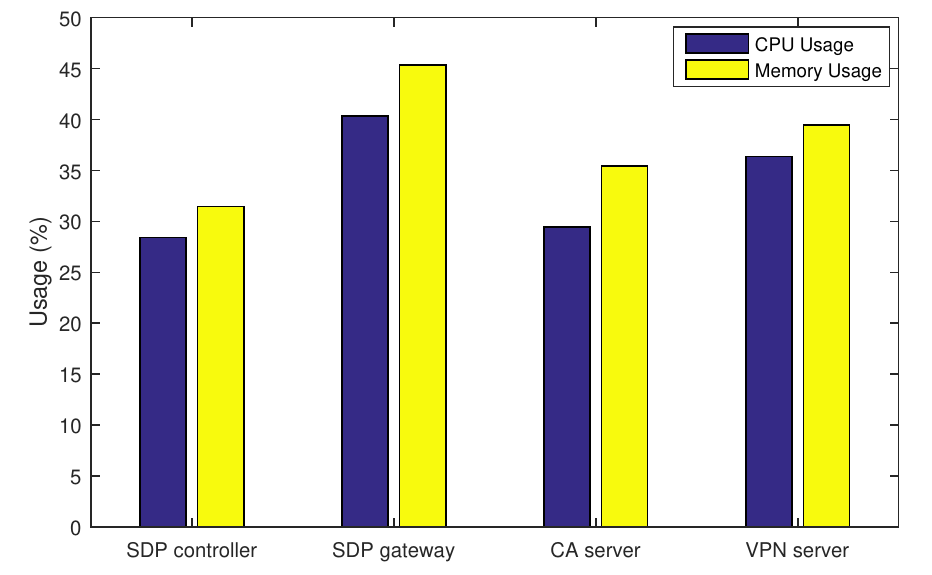}
    \caption{CPU and Memory utilization Comparison under SDP and VPN}
    \label{fig:CPUSDPVPN}
\end{figure}

\begin{figure}[h]
    \centering
    \includegraphics[width=0.5\textwidth]{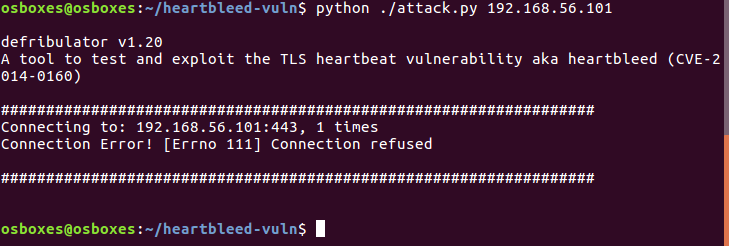}
    \caption{Heartbleed attack With SDP configurations}
    \label{fig:scan1}

\end{figure}

A port scanning attack was conducted on both SDP and VPN to evaluate their resistance against such attacks. The Nmap utility tool was used, and results are presented in Table II for SDP and VPN, respectively. When SDP framework was active, the attack targeted the gateway's ports 0-999, and all scanned ports were found to be closed. This demonstrates SDP gateway's ability to block unauthorized clients from accessing any ports within the network. Even if the attacker obtains the port numbers through a breach, network services cannot be accessed without a legitimate SPA. In contrast, when the same attack was performed on VPN (specifically AMF+SMF server, with the same result applicable to the UPF server), port 22 was found to be open, indicating the presence of an SSH service. This implies that an attacker could perform a sniffing attack to discover the ports for accessing the AMF+SMF server. Based on these results, the SDP framework exhibits superiority over VPN.

Figure~\ref{fig:CPUSDPVPN} illustrates CPU and memory utilization of SDP and VPN components. CPU and memory usage were measured in the VMs hosting the SDP controller, SDP gateway, CA server, and VPN server for a fair comparison. The SDP gateway requires higher CPU and RAM resources compared to the VPN server counterpart. This is due to additional tasks performed by the SDP gateway, such as forwarding SPA packets to the controller for verification and updating dynamic FWNKOP rules. The CA server exhibits slightly higher CPU and memory consumption compared to SDP controller because it carries out more tasks in the authentication process. Overall, both SDP and VPN utilize reasonable computing resources that are acceptable for most applications.

Figure~\ref{fig:scan1}, the result of a heartbleed attack on the SDP configuration is shown. The attack was targeted at SDP gateway 1 immediately after authentication with the SDP controller, aiming to steal private keys or other valuable server information. Despite SDP gateway 1 using OpenSSL version 1.0.1f, the connection was refused. This is because the SDP gateway only accepts valid SPA packets and is not vulnerable to heartbleed attack. It should be noted that the OpenVPN used in our implementation employs OpenSSL version 1.1.1 and is also immune to heartbleed attack. However, if OpenVPN utilizes older vulnerable versions of OpenSSL, a heartbleed attack can succeed.

\section{Discussion and Open Research Directions}

This paper proposes a software-based security solution for 6GC utilizing the SDP framework. A zero-trust architecture is implemented and evaluated against cyber attacks. Deploying SDP components with 5G NR and EPC network to establish a zero-trust environment, securing communications within and between them. OpenVPN is also implemented for comparison, but results demonstrate superiority of SDP in terms of protecting the entire service from unauthorized users. While OpenVPN exhibits lower resource utilization and initialization time, SDP remains within acceptable levels and exhibits higher resilience against port scanning attacks.

This work highlights key future research paths. Below, we delve into some of these directions.
\begin{itemize}
\item \textbf{Static networks related security challenges}: The static nature of systems and defenses significantly contributes to imbalanced security in cyberspace. Despite progress made in Artificial Intelligence and Machine Learning techniques for anomaly detection, static networks remain vulnerable and attractive targets for attackers. Moving target techniques aim to transform this nature of computer systems, effectively raising difficulty and cost associated with launching successful attacks. By implementing these techniques, systems become dynamic and elusive, making it challenging for cyber adversaries to target and exploit them 
\cite{x17}.

\item \textbf{Security challenges of the possible 6G RAN-Core convergence }: In the 6G architectural paradigm, 6G RAN functionalities and Core functionalities are expected to be combined to some extent. Some of the core functions are already virtual and distributive in nature in order to be implemented closer to the RAN to facilitate low-latency applications, while high-level RAN functions are being centralized. Combining the RAN and Core of the 6G to some extent can simplify the network and thus, pave the way for implementing more sophisticated services. However, this raises other security and privacy challenges, which require further research within the 6G paradigm \cite{9426946}.


\item \textbf{Security challenges related to the enabling technologies for 6G network}: The recent advancement of some cutting-edge technologies such as Artificial Intelligence (AI) techniques, Machine Learning (ML) techniques, Distributed Ledger Technology (DLI) such as Blockchain, Digital twin, intelligent edge computing are expected to facilitate the evolution to 6G networks. For example, the envisioned Intelligent Radio (IR) for 6G networks is expected to leverage AI/ML techniques to improve accurate channel modeling, resource allocation, beamforming, etc \cite{10115274}. However, these promising benefits come at the expense of an increase in security vulnerabilities, which are inherited from these enabling technologies. Therefore, more research efforts on security and privacy relating to these technologies are required to fully take advantage of the envisioned 6G networks.
\end{itemize}

\bibliographystyle{IEEEtran}
\bibliography{references}
\end{document}